\documentclass{llncs}
\usepackage[utf8]{inputenc}

\title{Pattern-based Subterm Selection in Isabelle\thanks{Presented at Isabelle Workshop 2014 in Vienna}}
\author{Lars Noschinski \and Christoph Traut}
\institute{Institut für Informatik, Technische Universität München, Germany\thanks{At the time of writing this article}}

\usepackage{ifthen}

\usepackage{palatino}

\usepackage{pifont}

\usepackage{colortbl}

\usepackage{amsmath}
\usepackage{amsfonts}
\usepackage{amssymb}
\usepackage{textcomp}
\usepackage{proof}

\usepackage{float}

\usepackage[english]{babel}
\usepackage{todonotes}
\usepackage{fixltx2e}
\usepackage[T1]{fontenc}
\usepackage[sc]{mathpazo}
\usepackage{listings}
\usepackage{stmaryrd}
\usepackage{isabelle}
\usepackage{isabellesym}
\usepackage{subcaption}
\usepackage{url}
\usepackage{hyperref}

\newcommand{\biglinebreak}{\vspace{2mm}\newline}

\newcommand*\justify{%
  \hyphenchar\font=`\-
}

\newcommand{\pattern}[1]{\texttt{\justify #1}}

\definecolor{matchcolor}{rgb}{0,0,0}

\newcommand{\match}[1] {\textcolor{matchcolor}{\boxed{#1}}}
\newcommand{\matchall}[1] {\underline{\match{#1}}}
\newcommand{\mmatch}[1] {\textcolor{matchcolor}{\boxed{#1}}}
\newcommand{\mmatchall}[1] {\underline{\mmatch{#1}}}

\newcommand{\pkw}[1]{\texttt{#1}}
\newcommand{\pat}{\pkw{at }}
\newcommand{\pin}{\pkw{in }}
\newcommand{\pfor}{\pkw{for }}

\newcommand{\pos}{\mathit{Pos}}
\newcommand{\Sem}[1]{\llbracket{#1}\rrbracket}
\newcommand{\sem}[1]{\mathcal{S}_\mathsf{#1}}
\newcommand{\Sb}{\sem{bind}}
\newcommand{\Si}{\sem{in}}
\newcommand{\Sp}{\sem{pos}}
\newcommand{\St}{\sem{term}}
\newcommand{\Sm}{\sem{match}}
\newcommand{\Sc}{\sem{concl}}
\newcommand{\Sa}{\sem{asm}}
\newcommand{\Sf}{\sem{for}}
\newcommand{\hole}{\Box}

\newcommand{\holepos}{\mathit{pos}_\hole}
\newcommand{\B}{\mathsf{B}}

\newcommand{\T}{\mathcal{T}}
\newcommand{\V}{\mathcal{V}}
\newcommand{\Lo}{\mathsf{L}}

\lstnewenvironment{mlpattern}{\lstset{ %
  backgroundcolor=\color{white},   
  basicstyle=\ttfamily,        
  breakatwhitespace=true,          
  breaklines=true,                 
  captionpos=b,                    
  commentstyle=\color{mygreen},    
  deletekeywords={...},            
  escapeinside={\%*}{*)},          
  extendedchars=true,              
  keepspaces=false,                
  keywordstyle=\color{black},     
  language=ML,                     
  morekeywords={*,...,definition, lemma, at, where, apply, fixes, shows, by, for},            
  numbers=none,                    
  numbersep=5pt,                   
  numberstyle=\tiny\color{mygray}, 
  rulecolor=\color{black},         
  showspaces=false,                
  showstringspaces=false,          
  showtabs=false,                  
  stepnumber=2,                    
  tabsize=2,                       
}}{}

\newcommand{\var}{?}

\begin{document}

\maketitle

\begin{abstract}
This article presents a pattern-based language designed to select (a set of) subterms of
a given term in a concise and robust way. Building on this language, we
implement a single-step rewriting tactic in the Isabelle theorem prover, which
removes the need for obscure "occurrence numbers" for subterm selection.

The language was inspired by the \emph{language of patterns} of Gonthier and
Tassi, but provides an elegant way of handling bound variables and a modular
semantics.

\end{abstract}

\section{Introduction}
\emph{Proof assistants}, sometimes called \emph{interactive theorem provers}, are tools designed to assist human users in the process of writing formal mathematical proofs. They allow users to write their proofs in a formal language, where every step of the proof is verified by the assistant and checked for logical errors. Two well-known proof assistants are \emph{Isabelle} and \emph{Coq}~\cite{17POTW}.

A common proof method is equational reasoning, either rewriting a term into a normal form or a manual application of equations. The latter is often used for proof exploration or when automated methods do not yield the desired result. As a given equation can often be applied to different subterms of a term, the user needs a way to select the desired subterm.
There are two obvious approaches to subterm selection which are implemented both in Coq and Isabelle.
As an example, consider the equation
\begin{align*}
 (a + b) + (c + d) = e + f
\end{align*}
and the commutativity rule \pattern{add\_commute}: $\var x + \var y = \var y + \var x$ (we use $\var$ to mark universally quantified variables). There are four positions where the commutativity rule can be applied. If we want to rewrite the subterm $c + d$ to $d + c$, current Isabelle versions offer basically three options:

\begin{itemize}
    \item \pattern{subst add\_commute[of "c" "d"]}
    \item \pattern{subst add\_commute[where x="c" and y="d"]}
    \item \pattern{subst (3) add\_commute}
\end{itemize}

The first two examples instantiate the variables in the rule, so that it is only applicable for the $c + d$ subterm. The third example instructs Isabelle to skip the first two subterms matching $\var x + \var y$ and rewrite the third (which happens to be $c + d $).

Instantiation can only partially resolve the ambiguities. Also, instead of specifying directly what to rewrite, the user needs to know the names or the order in which the variables occur in the rule. On the other hand, occurrence numbers are hard to read, as it is often not obvious which subterm has which occurrence number.

As discussed by Gonthier and Tassi~\cite{ALOPFSS}, both of these approaches are insufficient for writing maintainable proofs: While they are expressive enough to select any subterm, they lack the expressiveness for capturing the developers intent. As a result, they break easily when unrelated parts of the proof change. Gonthier and Tassi developed a \emph{language of patterns}~\cite{ALOPFSS} as an alternative approach. The idea is that (for rewriting) subterm selection already is done by a pattern: The left hand side of the rule. Instead of restricting this indirectly (by instantiating the rule), the pattern is given explicitly. If this does not suffice to unambiguously select a subterm, an additional context pattern can be used.

In this article, we present a new language for subterm-selection building upon the ideas of Gonthier and Tassi. In particular, our language adds the ability to match on bound variables in a robust way and provides an easy to understand and flexible syntax with formal (though not formalized) semantics. We implemented this language in the Isabelle theorem prover\footnote{\url{https://www21.in.tum.de/~noschinl/Pattern-2014/}}.

This article starts by introducing basic notation in Section \ref{preliminaries}. In Section \ref{combinators} we present a set of combinators for subterm selection and describe how to use them for rewriting in Section \ref{rewriting}. Section \ref{applications} shows some improvements on real-world examples before we conclude with Section \ref{conclusion}.

\section{Preliminaries}
\label{preliminaries}
Isabelle's term language is the language of the simply typed lambda calculus, where
variables bound by lambda abstractions are represented by de Bruijn
indices~\cite{DEBRUIJN}. In this article, we use the following simplified and
untyped term structure when talking about Isabelle terms.  For an infinite set
of variable names $\V$, the set of Isabelle terms $\T$ is inductively defined
as follows:
\begin{align*}
    &\V \subseteq \T
    &
    x \in \V \land t \in \T &\implies \Lambda_x t \in \T
    \\
    t_1, t_2 \in \T &\implies t_1 \; t_2 \in \T
    &
    n \in \mathbb{N}_0 &\implies \B_n \in \T
\end{align*}
$\V$ are the free variables and juxtaposition is application. The variable name
$x$ at the lambda abstraction $\Lambda_x t$ carries no semantics, but serves as
a note to the user. Two terms differing only in the variable attached to the $\Lambda$ symbols are considered equal ($\alpha$-equivalence). The \emph{bound variable} $\B_n$ refers to the $n+1$-th
lambda above it, i.e. in $\Lambda_x (f (\Lambda_y \B_1) \B_0)$ both bound
variables are bound by the outermost abstraction.  If $n$ is greater than the
number of abstractions above it, $\B_n$ is called a \emph{loose bound
variable}.

As in Isabelle, we usually hide the de Bruijn indices in our presentation. For
example, the term $\Lambda_x (f (\Lambda_y \B_1) \B_0)$ will be written as
$\lambda x.\; f\; (\lambda y.\; x)\; x$.
If a term contains loose bounds, we
index them by the number of missing abstractions:
\begin{align*}
    f\;\B_1 (\lambda c. \; \B_2)\;\B_0
    \text{ corresponds to }
    f\;\Lo_2 (\lambda c.\; \Lo_3)\;\Lo_1
\end{align*}
By $\#\Lo(t)$ we denote the number of different loose bound variables in the term $t$.

A pattern is a term which might also include the special wildcard symbol $\_$. A term \emph{matches} a pattern, if there are terms, such that replacing the wildcards by these terms, term and pattern become equal.

A position is a word over $\Sigma = \{a,l,r\}$, describing a subterm of a term.
The empty word $\varepsilon$ refers to the term itself, $a$ to the term under
an abstraction, and $l$ and $r$ to the left resp. right term of an application.
We write $\pos(t)$ for the set of positions of a term $t$, $pq$ for the
concatenation of positions $p$ and $q$, and $t|_p$ for the subterm of $t$ at
position $p$.

\section{Combinators for Subterm Selection}
\label{combinators}

If we abstract from rewriting, our problem can be phrased as follows: Given a
term, find a language to select the desired subterms (or rather,
their positions). The language should be expressive enough to select any
singleton set. It should allow for a concise description and be robust.

To this end, we describe a set of combinators, operating on sets of
term/position pairs, i.e., functions of the type
$\mathcal{P}(\T \times \Sigma^*) \to \mathcal{P}(\T \times \Sigma^*)$.

When giving examples, we write for example $S \simeq \mmatch{a + b} + c$ to denote that
the combinator function $S$ applied to $M = \{((a + b + c), \varepsilon)\}$
selected the subterm $a + b$ at the marked position. To denote that not only $a
+ b$, but also all its subterms were selected, we write $S \simeq {\mmatchall{a
+ b} + c}$ instead.

A rewrite rule can only be applied at a position which matches its left-hand
side $t$, so any rewriting method implicitly works as a filter, selecting only
the positions where the rule can be applied. This is expressed by the following
combinator:
\begin{definition}[Term combinator]
\begin{align*}
    \St(t) = M \mapsto \{(s, p) \mid (s,p) \in M \land s \text{ matches } t\}
\end{align*}
\end{definition}
As a second combinator, we introduce a way of selecting all subterms.
\begin{definition}[Subterm combinator]
\begin{align*}
    \Si = M \mapsto \{(s|_q, pq) \mid (s,p) \in M \land q \in \pos(s) \}
\end{align*}
\end{definition}
Now, composing these combinators gives us a way of selecting all positions
below a certain subterm:
\begin{example}
  \begin{align*}
      \St (\_ * \_)
      &\simeq
      (a + b) * c + d
      \\
      \Si
      &\simeq
      \mmatchall{(a + b) * c + d}
      \\
      \St (\_ * \_) \circ \Si
      &\simeq
      \mmatch{(a + b) * c} + d
      \\
      \Si \circ \St (\_ * \_) \circ \Si
      &\simeq
      \mmatchall{(a + b) * c} + d
      \\
      \St(a + \_) \circ \Si \circ \St(\_ * \_) \circ \Si
      &\simeq
      \mmatch{(a + b)} * c + d
  \end{align*}
\end{example}
Unfortunately, these combinator do not yet allow us a unique selection in every
case. For example, we are still missing a combinator $S$ such that $S \simeq
\match{a} = a$. One way to deal with this is to specify the position of the
left $a$ directly, i.e., as $l r$ (remember that $a = a$ corresponds to $(=
a)\; a$, when writing $=$ not as an infix).
\begin{definition}[Position combinator]
\begin{align*}
    \Sp(q) = M \mapsto \{(s|_q, pq) \mid (s,p) \in M \land q \in \pos(s) \}
\end{align*}
\end{definition}
Specifying the offset as a position is not very readable, so we mark the
position we are interested in by a special symbol in the pattern: If a pattern
$t$ contains a single hole symbol $\hole$, then $\holepos(t)$ is the position
of the hole in the term, i.e., $\holepos(\hole = \_) = l r$. Similar to $\_$\,,
it matches every term.

Consider the following term. How can we select the subterm $f\; b$?
\begin{align}
    \label{term_bvar}
    \lambda a\; b.\; [f\; a, (\lambda c.\; a), f\; b]
\end{align}
A possible solution would be to select the second element of this list, but
this might not capture our intent (and might be very cumbersome for larger
terms).
The obvious combinator $\St(f\;b)$ does not work, as $a$ and $b$ are free
variables in the pattern and bound variables in the term. The deeper problem is
that bound variables only exist as soon as we descend into the term. There are
various reasons why guessing a bound variable based on its name is not a viable
alternative:
Different bound variables with the same name might occur either in parallel
(e.g., $(\lambda x.\; x) \circ (\lambda x.\; x)$) or nested (e.g., $\lambda x.\; f\;
(\lambda x.\; x)$). Moreover, the terms from which we want to select subterms
are usually not input by the user directly, but are the result of some earlier
proof steps. As Isabelle considers terms modulo $\alpha$-equivalence, this
means the names might change in unexpected ways.

Hence, to refer to bound variables, we introduce new names when we descend into
a term. The $\St$ and $\Sp$ combinators allow us to control where to descend:
\begin{example}
  \label{ex:pos_term}
  \begin{align*}
      \Sp(\holepos(\lambda x\; y.\;\hole)) \circ \St(\lambda x\; y.\;\hole)
      \simeq \lambda a\; b.\; \mmatch{[f\;a, (\lambda c.\; a), f\;b]}
  \end{align*}
  The selected term is $[f\;\Lo_2, (\lambda c.\; \Lo_2), f\;\Lo_1]$.
\end{example}
If we replace the bound variables $\Lo_1$ and $\Lo_2$ by free variables, we can
refer to them in the $\St$ combinator. The $\Sb$ combinator introduces free variables.
\begin{definition}[Binding bound variables]
    The bind combinator $\Sb$ will
    replace the innermost loose bound variables by fresh variables:
    \begin{align*}
        \begin{split}
        \Sb(v_1, \dotsc, v_n) = M \mapsto \{ (s',p) \mid\;
            & (s,p) \in M \land n \leq \#\Lo(s)
            \\ &\land s' = \mathit{subst_\Lo}(s, [v_1, \dotsc, v_n])
        \}
        \end{split}
    \end{align*}
    where $\mathit{subst_\Lo}$ replaces $\Lo_1$ by $v_n$, $\Lo_2$ by $v_{n-1}$, \ldots, $\Lo_{n-1}$ by $v_2$
    and $\Lo_n$ by $v_1$.
\end{definition}
\begin{example}
    For the term selected in Example~\ref{ex:pos_term}, this works as follows:
    \begin{align*}
        \Sb(x,y) &\text{ selects }  [f\;x, (\lambda c.\; x), f\;y]
        \\
        \Sb(x) &\text{ selects }  [f\;\Lo_2, (\lambda c.\; \Lo_2), f\;x]
        \\
        \Sb(x,y,z ) &\text{ selects nothing}
    \end{align*}
\end{example}
So the following function lets us finally select the desired subterm $f\;b$ of Equation~\ref{term_bvar}:
\begin{align*}
    \St(f\;y)
    \circ \Sb(x,y)
    \circ \Sp(\holepos(\lambda x\; y.\;\hole))
    \circ \St(\lambda x\; y.\;\hole)
\end{align*}
Note that we used the same pattern both for $\Sp$ and $\St$ and that the use of
$\Sb$ only gives predictable results if we know where in the term we actually
are. Usually, this is the case after selecting with both $\St$ and $\Sp$ first.
So we combine the three combinators into a single one.
\begin{definition}[Matching combinator]
    The matching combinator selects the subterms according to a pattern $t$. If $t$ contains
    a single hole, it descends to the position of the hole and replaces all bound
    variables, which become loose in the descent, by free variables named like the corresponding bound variables in $t$.
    \begin{align*}
        \Sm(t) = \Sb(\mathit{binds}(t, \holepos(t))) \circ \Sp (\holepos(t)) \circ \St(t)
    \end{align*}
    By $\mathit{binds}(t,p)$ we denote the list of variables which are bound
    between the root and $p$ in $t$. E.g., for $t = \lambda a\;b.\; f\;a\;b$
    and $p = \holepos(\lambda a\;b.\; \hole)$, we have $\mathit{binds}(t,p) = (a,b)$.
\end{definition}
The matching combinator takes the variables it introduces directly from the
lambda abstractions. As this term was explicitly entered by the user, the
variables were not subject to renamings and produce the expected result.  When
using the bind combinator, one should take care of always using fresh variables
(otherwise, different variables in the term might be identified). With the
usual Isabelle naming mechanisms we will be able to hide this detail from the user.

We have described a basic set of combinators for navigating in a term. In
Isabelle, the subgoals we are rewriting usually have a special structure:
\begin{align*}
    \isasymAnd x_1\;x_2 \ldots x_m.\; P_1 \implies P_2 \implies \ldots \implies P_n \implies Q
\end{align*}
Here, the $x_i$ are universally quantified variables, the $P_i$ are premises
and $Q$ is the conclusion. When new universally quantified variables are
introduced during a proof, they are added at the end of the list.
This motivates the introduction of combinators for selecting these parts of a
goal.

\begin{definition}[Combinators for the goal structure]
For a term $t$ of the structure given above, $\Sc$ will select the term
$Q$ and $\Sa$ will select the terms $P_1, \ldots, P_n$ (with the appropriate
positions). For matching the universally quantified variables, it is important
to remember that the list is extended from the right, so usually one wants to
talk about a suffix of this list (cf. Isabelle's \texttt{rename\_tac}). So, if
$k \leq m$, $\Sf(y_1,\cdots,y_k)$ binds (as in $\Sb$) $x_m$ to $y_k$, $x_{m-1}$
to $y_{k-1}$, and $x_{m-k+1}$ to $y_1$ and selects the term
$P_1 \implies P_2 \implies \ldots \implies P_n \implies Q$.
\end{definition}

\begin{example}
    We demonstrate the $\Sf$ combinator:
    \begin{align*}
        \Sf(a,b) \simeq \isasymAnd x_1 x_2 x_3.\; \match{x_1 \leq x_2 \implies x_2 \leq x_3 \implies x_1 \leq x_3}
    \end{align*}
    In the matched term, the bound variables $x_2$ and $x_3$ are replaced by
    free variables $a$ and $b$. $x_1$ remains a loose bound variable:
    \begin{align*}
        \Lo_3 \leq a \implies a \leq b \implies \Lo_3 \leq b
    \end{align*}
    The bound variables $a$ and $b$ can then be used with the $\Sm$ combinator:
    \begin{gather*}
         \St(a \le b) \circ \Si \circ \Sf(a, b)
         \\
         \simeq
         \isasymAnd x_1\;x_2\;x_3.\;\; x_1 \le x_2 \implies \mmatch{x_2 \le x_3} \implies x_1 \le x_3
    \end{gather*}
\end{example}

\begin{example}
    The difference between $\Sf(a,b)$ and $\Sm(\isasymAnd a\; b.\; \hole)$ is that
    the former matches the rightmost and the latter the leftmost of the
    universally quantified variables (note that $\isasymAnd a\; b.\; t$ is a
    shorthand for $\isasymAnd a.\; \isasymAnd b.\; t$).
    \begin{align*}
        \Sf(a,b) &\simeq \isasymAnd x\;y\;z.\; \match{t\;x\;y\;z}
        \\
        \Sm(\isasymAnd a\;b.\; \hole) &\simeq \isasymAnd x\;y\; \match{\isasymAnd z.\; t\;x\;y\;z}
    \end{align*}
    where the selected term for the former is $t\;\Lo_3\;a\;b$ and the selected for the latter is $\isasymAnd z.\; t\;a\;b\;z$.
\end{example}

\section{Rewriting}
\label{rewriting}
In the last section we introduced a combinator language for subterm selection. We now apply this language to the implementation of a proof method \pattern{patsubst} for single-step rewriting.
We start by introducing a user-facing syntax for subterm selection, building on the previously introduced combinators.
\begin{mlpattern}
  <atom>    ::= <term> | concl | asm | prop
  <pattern> ::= (in <atom> | at <atom> | for <names>)
                  [<pattern>]
\end{mlpattern}
The atom \texttt{<term>} is parsed into a term $t$.
We define the semantics of atoms as follows:
\begin{align*}
    \Sem{t} &= \Sm(t)
    &
    \Sem{\texttt{concl}} &= \Sc
    &
    \Sem{\texttt{asm}} &= \Sa
    &
    \Sem{\texttt{prop}} &= \mathit{I}
\end{align*}
Based on this, the semantics of patterns are:
\begin{align*}
    \Sem{\texttt{in}\;a} &= \Si \circ \Sem{a} \\
    \Sem{\texttt{at}\;a} &= \Sem{a} \\
    \Sem{\texttt{\pfor} \; (i_1 \ldots i_n)} &= \Sf(i_1, \ldots, i_n) \\
    \Sem{\mathit{pat}_1 \; \mathit{pat}_2} &= \Sem{\mathit{pat}_1} \circ \Sem{\mathit{pat}_2}
\end{align*}

To select subterms of some term $t$ with a pattern $p$, $\Sem{p}$ is applied to an initial set $M(t)$ to obtain the set of selected subterms $\Sem{p}(M(t))$. As in our example syntax, we set $M(t) = \{(t, \varepsilon)\}$.

This allows full control over the subterm selection. From our experience with the existing \pattern{subst} tactic for single-step rewriting in Isabelle, rewriting in the conclusion is the most common case. For this reason, our implementation automatically appends \pattern{\pin concl} if a pattern ends with a term atom.

\begin{example}
    \mbox{}\\
\setlength\extrarowheight{2mm}
\begin{tabular}{l c l}
  \pattern{\pat "a + \_"} &
    $\simeq$ &
    $(\match{a + b}) + c$ \\
  \pattern{\pat "a + \_" \pat prop} &
    $\simeq$ &
    $(a + b) + c$ \\
  \pattern{\pin "\_ * c"} &
    $\simeq$ &
    $\matchall{(a + b) * c} + d$ \\
  \pattern{\pat "\_ + \_" \pin "\_ * c"} &
    $\simeq$ &
    $(\match{a + b}) * c + d$ \\
  \pattern{\pat "x + 1" \pin "f \_"} &
    $\simeq$ &
    $f\; (\match{x + 1}) = g\; (x + 1)$ \\
  \pattern{\pat "$\hole$ = \_"} &
    $\simeq$ &
    $\match{x} = x$ \\
  \pattern{\pat "a + \_" \pin "f $\hole$ + f \_"} &
    $\simeq$ &
    $f\; ((\match{a+b})+c) + f\; ((a+b)+c)$ \\
  \pattern{\pat "v+1" \pin "($\lambda$v.$\;\hole$) $\equiv$ \_"} &
    $\simeq$ &
    $(\lambda x.\;(\match{x+1})+c) \equiv (\lambda y.\; (y+1)+c)$ \\
\end{tabular}
\end{example}

\subsection{Conversions}
For a long time, Isabelle has provided the \pattern{subst} proof method for
single-step rewriting. This proof method uses the rule
\begin{align*}
    \frac{f \equiv g\qquad x \equiv y}{f\;x \equiv g\;y}
\end{align*}
to rewrite a term in a single derivation step: $x$ is chosen as a subterm matching the left hand side of the applied equation, $f$ as the surrounding context. Then the equation is applied to $x$, resulting in $y$, and $g$ is the same as $f$.
Unfortunately, this approach is incompatible with our handling of
bound variables.
Instead, our implementation builds upon Isabelle's \emph{conversions}.

A conversion is a function that takes a term $t$ and produces a theorem of form $t \equiv t'$. Apart from the basic conversion which rewrites a term at its root, the combinators \pattern{abs\_conv}, \pattern{arg\_conv}, and \pattern{fun\_conv} are of prime interest to us: These combinators take a conversion and apply it to the term under an abstraction, the argument part of an application, and the function part of an application, respectively. Figure~\ref{fig:conversions} depicts the rules implemented by these conversions. When descending below the abstraction, the \pattern{abs\_conv} conversion automatically invents a fresh free variable, based on the name of the bound variable. A small modification to this function and delayed parsing of the pattern terms ensure that the name presented to the user is exactly the name of the bound variable (without introducing variable clashes).

\begin{figure}
    \begin{center}
        \begin{subfigure}[c]{.33\linewidth}
            \begin{align*}
                \frac{t_1[x/s] \equiv t_2[x/s]}{(\lambda s. \; t_1) \equiv (\lambda s. \; t_2)}
            \end{align*}
            \caption{\pattern{abs\_conv}}
            \label{abs_conv}
        \end{subfigure}%
        \begin{subfigure}[c]{.33\linewidth}
            \begin{align*}
                \infer{f\, s \equiv f\, t}{s \equiv t}
            \end{align*}
            \caption{\pattern{arg\_conv}}
        \end{subfigure}%
        \begin{subfigure}[c]{.33\linewidth}
            \begin{align*}
                \infer{f\, s \equiv g\, s}{f \equiv g}
            \end{align*}
            \caption{\pattern{fun\_conv}}
        \end{subfigure}%
    \end{center}
    \caption{
        Rules implemented by the conversion combinators. By $t[x/s]$ we denote substitution of $x$ for $s$ in $t$ . In (\ref{abs_conv}), $x$ must not be free in $t_1$ and $t_2$.
    }
    \label{fig:conversions}
\end{figure}

Note that the conversions from Figure~\ref{fig:conversions} functions directly relate to subterm positions:
\pattern{fun\_conv} corresponds to $l$, \pattern{arg\_conv} to $r$, and \pattern{abs\_conv} to $a$.
 The position $\epsilon$ can be represented by the identity function $I$.
Also, concatenation of positions corresponds to function composition.

In our semantics, we represent subterms by pairs of term and position. In our implementation, we represent each of these positions by a conversion combinator, i.e., a function of type $\mathit{conv} \rightarrow \mathit{conv}$, where $\mathit{conv}$ is the type of conversions.
During subterm selection, we descend into terms, while updating the current position by composing the appropriate function with it.
This implicit representation of the position makes for a straight-forward implementation and avoids having to descend into the term multiple times.

In contrast to the old \pattern{subst} implementation, Isabelle's conversions cannot deal with conditional equations.
Support for conditional equations is an important feature, in particular for proof exploration. Fortunately, this problem is not inherent to our approach and can be rectified by using a more general variant of conversions.

\paragraph{Manual instantiation}
There are rewrite rules that introduce a new variable, one simple example being $0 = 0 * ?a$. Another example would be rules for adding an invariant annotation to a loop in program verification. In Example \ref{ex:while}, we will show a simple example of such a rule.

Such a variable will not be instantiated automatically (as there is no indication what its value should be), so usually we need to instantiate it manually. If its value does not depend on any bound variable, the variables can be instantiated before rewriting. If the value depends on a bound variable, this is not possible; the complications being the same as for subterm selection. Currently, Isabelle provides no easy way around this problem.

However, in our implementation, the use of \pattern{abs\_conv} ensures that all bound variables under which we descended were replaced by free variables. So, at the point where we apply the rewriting, we will never need to instantiate the term with any bound variables, but with the associated free variables instead. The same technique that allows us to match bound variables now allows us to instantiate the theorem with bound variables.

Our syntax for instantiation during term rewriting mirrors the syntax of the \pattern{where} attribute, which can be used to instantiate theorems directly. The \pattern{where} keyword is followed by a list of instantiations, which are pairs of variable name and term.

We demonstrate the use of this feature on the \pattern{while} combinator from Isabelle's library.
\begin{example}
\label{ex:while}
The following program implements multiplications by addition, which in turn is implemented as repeated increment.
\begin{mlpattern}
    while (%*$\lambda$*)(i, x). i < a)
      (%*$\lambda$*)(i::nat, x::nat).
        let
          (_, x') = while (%*$\lambda$*)(j, _). j < b)
            (%*$\lambda$*)(j::nat, x). (j + 1, x + 1))
            (0, x)
        in (i + 1, x'))
      (0, 0)
\end{mlpattern}
Proofs about the while combinator usually use the relevant introduction rule, which requires specifying the loop invariant by hand. So a goal involving \pattern{while} needs to be decomposed by hand, before automated tools can be applied. By using an annotated variant of \pattern{while}, one can write an introduction rule picking up the invariant from the annotation.
\begin{mlpattern}
  definition
    "while_inv (I :: 'a %*$\Rightarrow$*) bool) c b s :: 'a %*$\equiv$*) while c b s"
\end{mlpattern}

Now, assume we wanted to add the invariant \pattern{$\lambda$(j, x).\,x = j + a * i} to the inner while loop, where the variable \pattern{i} is the variable bound in the body of the outer while loop. For this, we need to substitute \pattern{while\_inv} for \pattern{while} and instantiate the schematic variable \pattern{?I} with the invariant, using the following command:
\pagebreak[2]
\begin{mlpattern}
  apply (
    patsubst in "while _ (%*$\lambda$*)(i, _). %*$\hole$*)) _"
    while_inv_def[symmetric]
    where ?I = "%*$\lambda$*)(j, x). x = j + a*i"
  )
\end{mlpattern}
We use a hole pattern to select the inner while loop for rewriting and also associate the identifier \pattern{i} to the bound variable that should appear in our invariant. Then we instantiate the rewrite rule with our invariant. Since we associated the identifier \pattern{i} with a bound variable, any occurrence of \pattern{i} in the instantiation will be replaced with this particular bound variable.
\end{example}

The \pattern{where} option we just described requires the user to know the names of the variables in the rewrite rule, which is one of the objections we mentioned in the introduction. As an alternative, \pattern{patsubst} could take a second pattern, describing the expected term after rewriting.

As an additional benefit, such a change would allow selecting the correct rule from a set of rules: For example, we could state our intent of rewriting $b + a$ to $a + b$, together with a set of algebraic rules. Knowing both the left and the right hand side, the proof method can then select the symmetry rule and apply it.

\section{Practical applications}
\label{applications}
In this section, we look at some uses of \pattern{subst} in Isabelle's Archive
of Formal Proofs~\footnote{\url{http://afp.sourceforge.net}} and demonstrate that our new \pattern{patsubst}
method indeed allows for writing much nicer proofs.

\subsection{Real world usage of subst}
To gain some insight into the usage of \pattern{subst}, we decided to look at the \emph{Archive of Formal Proofs} (AFP), a large repository of formal proofs for Isabelle. We used the most recent version of the AFP, which, at the time of this writing, was from the 14th of January 2014.
\biglinebreak
It is difficult to make any general statement about the usage of \pattern{subst}, and the usefulness of our pattern-based implementation, from the data we collected, because there are several factors that may distort our results:
\begin{itemize}
 \item \pattern{subst} is often used during proof exploration. This is, of course, not visible in the final proof, since those exploratory statements naturally are removed as soon as a suitable proof is found.
 
 \item It is also possibility that some proof authors avoid using \pattern{subst} exactly because of the problems associated with subterm selection that this work intends to address. Even though these authors might benefit from our new \pattern{subst} implementation, we will not be able to see any evidence of this in the AFP.

 \item We searched the AFP using regular expressions to find usages of \pattern{subst}. Our results are not completely accurate, since we are limited by the expressiveness of regular languages. Since some of our results are only approximate, we will, in these cases, only give approximate figures.
\end{itemize}

\subsubsection*{Usage of \pattern{subst} in the AFP}
%
%
%
%
%

Our search revealed that the AFP contains an total of 1654 theory files (i.e. files with the extension \emph{*.thy}). Of those, only about 380 contain usages of \pattern{subst}, which is a little more than 20\%. In total, there are about 4000 distinct usages of \pattern{subst} in the AFP.

We found 135 usages of \pattern{subst} that also used the \pattern{where} attribute to instantiate the applied rule. Those usages where spread across 45 files. The \pattern{of} attribute appears to be the more popular choice when rewriting with \pattern{subst}, we found over 700 usages in about 60 files. We also counted 177 instances in about 50 files where \pattern{subst} was used with at least one occurrence number.

This makes for a total of roughly 1000 usages of \pattern{subst} in the AFP where our pattern-based approach could potentially add value.

\subsection{Examples}
We will now examine some real world usages of \pattern{subst}, and show how they can benefit from our pattern based implementation. All of these examples are taken from the AFP. They therefore were embedded inside relatively big theories, sometimes several thousand lines long. In this section, we only want to focus on \pattern{subst} itself, so we present these examples without any context.

We chose the following examples because they each use a different method for subterm selection and they all benefit from our pattern approach. But they most certainly are not isolated instances where our approach works. Due to the sheer volume of potential examples available, we easily could have chosen other examples that would have worked just as well.

\begin{example}[Cauchy/CauchysMeanTheorem.thy]
Even though our previous examples where all constructed, they were not totally removed from reality, as this example shows. Consider the following subgoal:
\begin{mlpattern}
  (len + 1) * %*$\Sigma$*):xs / (len * (1 + len)) = %*$\Sigma$*):xs / len
\end{mlpattern}
To this, we want to apply the rule \pattern{mult\_commute}:
\begin{mlpattern}
  ?a * ?b = ?b * ?a
\end{mlpattern}
For this, the original author used the following command:
\begin{mlpattern}
  apply (subst mult_commute [where a="len"])
\end{mlpattern}
Using our pattern language, we can replace this with the following command:
\begin{mlpattern}
  apply (patsubst at "len * _" mult_commute)
\end{mlpattern}
This admittedly is only a relatively minor improvement. It arguably improves readability, but most importantly, it removes the variable name \pattern{a} from the command. The proof author no longer has to know the names of the variables in the \pattern{mult\_commute} rule to apply it at the correct location.
\end{example}

\begin{example}[Group-Ring-Module/Algebra1.thy]
Here, the user applies the definition \pattern{segment\_def}:
\begin{mlpattern}
  segment ?D ?a =
  (if ?a %*$\notin$*) carrier ?D then carrier ?D
   else {x. x %*$ \prec $*) ?D ?a %*$ \wedge $*) x %*$ \in $*) carrier ?D})
\end{mlpattern}
To this proof state:
\begin{mlpattern}
  a %*$ \in $*) carrier D %*$\Longrightarrow$*)
  b %*$ \in $*) carrier D %*$\Longrightarrow$*)
  a %*$\prec$*) b %*$\Longrightarrow$*) segment (Iod D (segment D a)) b = segment D a
\end{mlpattern}
This is the command the user chose to accomplish this:
\begin{mlpattern}
  apply (subst segment_def[of "Iod D (segment D a)"])
\end{mlpattern}
Again, we can easily resolve the ambiguity without instantiating our rule:
\begin{mlpattern}
  apply (patsubst at "%*$\hole$*) = _" segment_def)
\end{mlpattern}
This command is not only far less verbose, it also explicitly states the author's intention to rewrite the outermost occurrence on the left hand side of the equation.
\end{example}

\begin{example}[Coinductive/Coinductive\_List.thy]
In this example, \pattern{subst} is used with an occurrence number to apply the rule \pattern{lmap\_ident[symmetric]}:
\begin{mlpattern}
  ?t = lmap (%*$\lambda$*)x. x) ?t
\end{mlpattern}
This rule is highly ambiguous. It can be applied to any subterm of the right
type. There are eight such subterms in the following term:
\begin{mlpattern}
  lzip (lmap f xs) ys =
    lmap (%*$\lambda$*)(x, y). (f x, y)) (lzip xs ys)
\end{mlpattern}
To disambiguate the application of the rule, the user chose to state an occurrence number.
\begin{mlpattern}
  apply (subst (4) lmap_ident[symmetric])
\end{mlpattern}
This command is completely unreadable. It is not obvious at all which subterm the user wishes to rewrite.
In this case, the user wanted to apply the rule to the variable \pattern{ys} on the left hand side of the equation. To make this intention clear, we propose using the following command:
\begin{mlpattern}
  apply (patsubst at ys in "%*$\hole$*) = _" lmap_ident[symmetric])
\end{mlpattern}
\end{example}

\section{Conclusion}
\label{conclusion}

In this article, we presented a combinator language for subterm selection and, building on this, implemented a single-step rewrite tactic for the Isabelle theorem prover. The purpose of single-step rewriting is to give the user full control over the rewriting process. In contrast to the existing \pattern{subst} proof method, our approach to subterm selection does support this spirit of explicitness. Patterns express the user's intention during rewriting; containing information both about the targeted subterm and its context. In addition to this, they enable us to explicitly instantiate with bound variables during rewriting. Previously, this was not possible with any rewriting tool.  Compared with the language of patterns by Gonthier and Tassi we contribute handling of bound variables and a more regular and modular semantics.

A very useful addition to the rewriting implementation presented here would be support for \emph{congruence rules}. Congruence rules provide additional assumptions when rewriting under certain contexts. As an example, when rewriting the branches of an conditional expression, we may assume the condition to hold respectively not hold:
\begin{align*}
    \infer
    {(\mathrm{if}\; b\; \mathrm{then}\; x\; \mathrm{else}\; y) = (\mathrm{if}\; c\; \mathrm{then}\; u\; \mathrm{else}\; v)}
    { b = c \qquad c \implies x = u \qquad \lnot c \implies y = v}
\end{align*}
These kind of rules are heavily used by the Isabelle's automatic simplifier.

It might also be interesting to support ``rewriting'' for weaker relations. Window inference\cite{grundy1991window} is a proof technique for program refinement with respect to transitive and reflexive relations. This technique allows ``opening a window'' at any position in the term and refine it according to the given relation. Our subterm selection language would allow a comfortable selection of this position.

\bibliographystyle{splncs03}
\bibliography{main}

\begin{thebibliography}{1}
\providecommand{\url}[1]{\texttt{#1}}
\providecommand{\urlprefix}{URL }

\bibitem{DEBRUIJN}
De~Bruijn, N.G.: Lambda calculus notation with nameless dummies, a tool for
  automatic formula manipulation, with application to the church-rosser
  theorem. In: Indagationes Mathematicae (Proceedings). vol.~75, pp. 381--392.
  Elsevier (1972)

\bibitem{ALOPFSS}
Gonthier, G., Tassi, E.: A language of patterns for subterm selection. In: ITP
  2012, pp. 361--376. Springer

\bibitem{grundy1991window}
Grundy, J.: Window inference in the {HOL} system. In: TPHOLs 1991. pp.
  177--189. IEEE (1991)

\bibitem{17POTW}
Wiedijk, F., Scott, D.S.: The seventeen provers of the world. Springer (2006)

\end{thebibliography}

\end{document}